\documentclass[preprint,aps,showpacs,nofootinbib]{revtex4}

\usepackage{graphicx}
\usepackage{epsfig,amssymb,amsmath,color}

\newcommand{\beq}{\begin{equation}}
\newcommand{\eeq}{\end{equation}}
\newcommand{\bea}{\begin{eqnarray}}
\newcommand{\eea}{\end{eqnarray}}
\newcommand{\bear}{\begin{array}}
\newcommand {\eear}{\end{array}}
\newcommand{\bef}{\begin{figure}}
\newcommand {\eef}{\end{figure}}
\newcommand{\bec}{\begin{center}}
\newcommand {\eec}{\end{center}}

\newcommand{\la}{\left\langle}
\newcommand{\ra}{\right\rangle}

\def\EQ#1{Eq.~(\ref{#1})}

\def\REF#1{(\ref{#1})}
\def\GEV#1{10^{#1}{\rm\,GeV}}

\def\lrfp#1#2#3{ \left(\frac{#1}{#2} \right)^{#3}}

\begin{document}
\draft
\tighten
\preprint{
KEK-TH-1711,
DESY-14-033,
TU-958,
IPMU14-0060
}
\title{\large \bf
Solving the Tension between High-Scale Inflation \\
and Axion Isocurvature Perturbations
}
\author{
    Tetsutaro Higaki\,$^{a,\star}$\footnote[0]{$^\star$ email: thigaki@post.kek.jp},
    Kwang Sik Jeong\,$^{b,\ast}$\footnote[0]{$^\ast$ email: kwangsik.jeong@desy.de},
    Fuminobu Takahashi\,$^{c,\,d\,\dagger} $\footnote[0]{$^\dagger$ email: fumi@tuhep.phys.tohoku.ac.jp}
    }
\affiliation{
    $^a$ Theory Center, KEK, 1-1 Oho, Tsukuba, Ibaraki 305-0801, Japan \\
    $^b$ Deutsches Elektronen Synchrotron DESY, Notkestrasse 85, 22607 Hamburg, Germany \\
    $^c$ Department of Physics, Tohoku University, Sendai 980-8578, Japan \\
    $^d$ Kavli IPMU, TODIAS, University of Tokyo, Kashiwa 277-8583, Japan
    }

\vspace{2cm}

\begin{abstract}
The BICEP2 experiment determined the Hubble parameter during inflation to be about $10^{14}$\,GeV.
Such high inflation scale is in tension with the QCD axion dark matter if the Peccei-Quinn (PQ)
symmetry remains broken during and after inflation, because too large  axion isocurvature
perturbations would be  generated. The axion isocurvature perturbations  can be suppressed if the axion acquires a sufficiently heavy
mass during inflation. We show that this is realized if the PQ symmetry is explicitly broken down to a discrete
symmetry and if the breaking is enhanced during inflation.
 We also show that, even when the PQ symmetry becomes spontaneously
 broken after inflation,  such a temporarily enhanced PQ symmetry breaking relaxes
 the constraint on the axion decay constant.
\end{abstract}
\maketitle


\section{Introduction}

The identity of dark matter is one of the central issues in cosmology and particle physics.
Among various candidates for dark matter, the QCD axion is a plausible and interesting
candidate.  The axion, $a$,  arises as a pseudo-Nambu-Goldstone (pNG) boson in association with the spontaneous
breakdown of a global U(1)$_{\rm PQ}$ Peccei-Quinn (PQ)  symmetry~\cite{Peccei:1977hh,QCD-axion}.
If the U(1)$_{\rm PQ}$ symmetry is explicitly broken only by the QCD anomaly,
the axion is stabilized at vacuum with a vanishing CP
phase, solving the strong CP problem. More important, the dynamical relaxation  necessarily induces coherent
oscillations of axions, which contribute to cold dark matter (CDM).
We focus on the axion  CDM
which accounts for the total dark matter density, throughout this letter.

The axion mass receives contributions from the QCD anomaly,
\bea
\label{maQCD}
m^{\rm QCD}_a \simeq 6\times 10^{-6}{\rm eV} \left(\frac{f_a}{10^{12}{\rm GeV}}\right)^{-1},
\eea
where $f_a$ is the axion decay constant. Because of the light mass, the axion  generically acquires quantum
fluctuations of  $\delta a \simeq H_{\rm inf}/2\pi$ during inflation, leading to CDM isocurvature perturbations. Here $H_{\rm inf}$
is the Hubble parameter during inflation.   The mixture of the isocurvature perturbations is tightly constrained by
the Planck observations~\cite{Ade:2013uln}, which reads
\bea
H_{\rm inf} < 0.87\times 10^7\,{\rm GeV}
\left(\frac{f_a}{10^{11}{\rm GeV}}\right)^{0.408}~~~(95\%\,{\rm CL}),
\label{iso}
\eea
neglecting anharmonic effects~\cite{Turner:1985si,Lyth:1991ub,Bae:2008ue,Visinelli:2009zm,Kobayashi:2013nva}.
In particular, a large-field inflation such as chaotic inflation~\cite{Linde:1983gd} is in conflict with the
isocurvature bound. Recently the BICEP2 experiment announced the discovery of the primordial B-mode
polarization~\cite{BICEP2}, which determines the inflation scale as
\bea
\label{B}
H_{\rm inf} &\simeq& 1.0 \times \GEV{14} \lrfp{r}{0.16}{\frac{1}{2}},\\
r &=& 0.20^{+0.07}_{-0.05} ~~(68\%{\rm CL}),
\eea
where $r$ denotes the tensor-to-scalar ratio.\footnote{Such large tensor-to-scalar ratio can be explained
in various large field inflation; see e.g.~\cite{Freese:1990ni,Kawasaki:2000yn,Silverstein:2008sg,
McAllister:2008hb,Kaloper:2008fb,Takahashi:2010ky,Nakayama:2010kt,Nakayama:2010sk,Harigaya:2012pg,Croon:2013ana,
Nakayama:2013jka,Czerny:2014wza,Czerny:2014xja,Nakayama:2014-HCI}. The tension with the Planck result can be
relaxed in the presence of small modulations in the inflaton potential~\cite{Kobayashi:2010pz} or hot dark matter/dark radiation. }
 After subtracting the best available estimate for foreground dust,
the allowed range is modified to $r = 0.16^{+0.06}_{-0.05}$.
Therefore one can see from \REF{iso} and \REF{B} that
there is a clear tension between the inflation scale determined by the BICEP2 and the QCD axion dark matter.\footnote{
The isocurvature perturbation bound similarly applies to the so called axion-like particles, or general pseudo Nambu-Goldstone
bosons, which are produced by the initial misalignment mechanism and contribute to dark matter.
}

There are various known ways to suppress the axion CDM isocurvature perturbations.
First, if the PQ symmetry is restored during inflation (or reheating),  there is no axion CDM isocurvature perturbations,
as  the axion appears only when the PQ symmetry is spontaneously broken some time after inflation~\cite{Linde:1990yj,Lyth:1992tx}.
In this case topological defects such as axionic cosmic strings and domain walls are generated,
and in particular the domain wall number $N_{\rm DW}$ must be unity to avoid the cosmological catastrophe~\cite{Hiramatsu:2012gg}.
Second, if the kinetic term coefficient for the phase of the PQ scalar was larger during inflation
than at present, the quantum fluctuations, $\delta a$, can be suppressed after inflation.
This is possible if the radial component
of the PQ scalar takes a larger value during inflation~\cite{Linde:1990yj,Linde:1991km}.
The scenario can be implemented easily in a supersymmetric (SUSY) theory, as
the saxion potential is relatively flat, lifted by SUSY breaking effects.
Interestingly, a similar effect is possible if there is a non-minimal coupling to gravity~\cite{Folkerts:2013tua}.
Third, the axion may acquire a heavy mass during inflation so that its quantum fluctuations get suppressed~\cite{Jeong:2013xta}.
In Ref.~\cite{Jeong:2013xta} two of the present authors (KSJ and FT) showed that
the QCD interactions become strong at an intermediate or high energy scale in the very early Universe, if the Higgs field has a
sufficiently large expectation value.\footnote{ The idea of heavy QCD axions during inflation was
considered in Refs.~\cite{Dvali:1995ce,Banks:1996ea,Choi:1996fs} to suppress the axion abundance, not the isocurvature
perturbations.}

In fact, the second solution of Refs.~\cite{Linde:1990yj,Linde:1991km} is only marginally
consistent with the BICEP2 result \REF{B}, if the field value of the PQ scalar is below the Planck scale.
Also the third solution of Ref.~\cite{Jeong:2013xta} is consistent with the BICEP2
result only in a corner of the parameter space. Therefore, we need another solution to suppress the axion isocurvature
perturbations, as long as we assume that the PQ symmetry remains broken during and after inflation.

In this letter, we propose a simple mechanism to suppress the axion CDM isocurvature perturbations
along the line of the third solution. Instead of making the QCD interactions strong during inflation, we introduce
a PQ symmetry breaking operator, which becomes relevant only during inflation. If the axion acquires a sufficiently
heavy mass during inflation, the axion CDM isocurvature perturbations practically vanish, evading the isocurvature
bound on the inflation scale. After inflation,
the explicit PQ breaking term should become sufficiently small so that it does not spoil the axion solution to the strong CP problem.

There are various possibilities to realize such temporal enhancement of the axion mass.
If, during inflation, the radial component
of the PQ scalar, i.e. the saxion, takes a large field value, the PQ symmetry breaking operator is enhanced and
the axion becomes heavy. After inflation, the saxion settles down at the low-energy minimum located at a smaller field
value where the explicit PQ breaking term
is sufficiently small. Alternatively we can consider a case in which  the PQ symmetry breaking operators are present only during inflation.
This is the case if the inflaton is coupled to the PQ symmetry breaking operators; during inflation, the operators are enhanced
due to a large vacuum expectation value (VEV) of the inflaton, whereas they are suppressed if the inflaton is stabilized
at much smaller field values after inflation.
 This can be nicely implemented
 in a large-field inflation model where the inflaton field value evolves significantly.

 Later in this letter we also briefly consider the first solution to the tension between the BICEP2 results and
  the axion CDM isocurvature perturbations, i.e., the PQ symmetry restoration during or after inflation.
 We will show that, even in this case, the temporarily enhanced PQ symmetry breaking relaxes the bound on the axion
 decay constant, allowing $f_a \gtrsim \GEV{10}$ and also $N_{\rm DW} \ne 1$.

\section{Isocurvature constraints on the axion CDM}

\begin{figure}[t!]
  \begin{center}
    \includegraphics[scale=0.8]{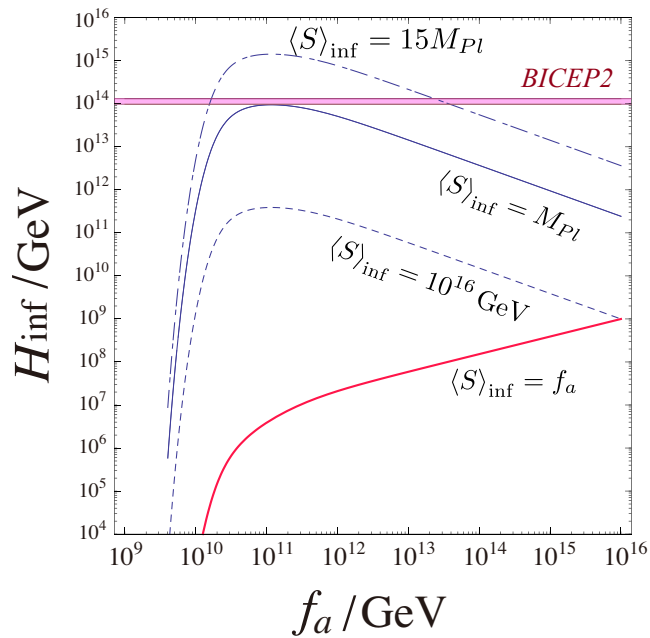}
  \end{center}
  \caption{Constraint on the inflation scale from the axion CDM isocurvature perturbation.
  The isocurvature constraint \REF{iso} is shown by the solid (orange) line, where the anharmonic effect is taken
  into account~\cite{Kobayashi:2013nva}.
  The relaxed constraints for $\langle S \rangle_{\rm inf} = \GEV{16}, M_{Pl}$ and $15 M_{Pl}$ are shown
  by the dashed, solid, and dot-dashed lines, respectively.
%
  The horizontal band is the BICEP2 result \REF{B}.
  If the axion acquires a heavy mass during inflation, all these constraints disappear as shown in the text.
}
  \label{fig:relaxed-constraint}
\end{figure}

The axion, if exists during inflation,  acquires quantum fluctuations, $\delta a = H_{\rm inf}/2\pi$,
giving rise to the axion CDM isocurvature perturbations.  The isocurvature constraint on the axion CDM
leads to the upper bound on the Hubble parameter during inflation as in \EQ{iso}, which is shown by
the solid (red) line Fig.~\ref{fig:relaxed-constraint}. Here the anharmonic effect is
taken into account~\cite{Kobayashi:2013nva}; the axion CDM isocurvature perturbations get significantly
enhanced as the initial field value approaches the hilltop, as can be seen for $f_a \lesssim \GEV{11}$.
Note that  we assume that the axion produced by the initial misalignment mechanism accounts for
the total CDM density in the figure.

Let us here briefly study how much the second solution mentioned in the Introduction can relax the isocurvature
bounds on the Hubble parameter. To see this, we write the kinetic term of the PQ scalar, $S$, whose expectation
value determines the  axion decay constant, as:
\bea
{\cal L}_K &=& \zeta^2 \partial S^\dag \partial S \supset \zeta^2 |S|^2 (\partial \theta)^2,
\eea
where we define $S = |S| e^{i \theta}$, and $\zeta (>0)$ parametrizes a possible deviation from
the canonically normalization. In general, $\zeta$ may depend on $S$, the inflaton or other fields: $\zeta = \zeta (\Phi)$,
where $\Phi$ denotes such fields collectively.
During inflation, the canonically normalized axion, $a = \la \zeta |S| \ra_{\rm inf} \theta$, acquires quantum fluctuations
of order $H_{\rm inf}$ at the horizon exit, namely,
\bea
\delta \theta &=& \frac{H_{\rm inf}}{2\pi \la \zeta |S| \ra_{\rm inf} }.
\eea
Note that the quantum fluctuations, $\delta \theta$,  at super-horizon scales remain constant, even if $\zeta^2 |S|^2$ evolves in time  during and after inflation.
Thus, the quantum fluctuations of the canonically normalized axion in the low energy is
given by
\bea
\delta a &=&  \frac{\la \zeta |S| \ra_0}{\la \zeta |S| \ra_{\rm inf}}  \frac{H_{\rm inf}}{2\pi},
\eea
where the subindices $0$ and ${\rm inf}$ denote that the variables are estimated in the low energy
and during inflation, respectively. For $\zeta = 1$, if $|S|$ takes a large VEV during inflation and settles
down at a smaller field value in the low energy, the axion quantum fluctuations are suppressed
by a factor of $\langle S \rangle_{\rm inf}/f_a$.\footnote{We note that the one cannot take $\langle S \rangle_{\rm inf}$
arbitrarily large, as it might affect the inflaton dynamics. In non-SUSY case, we need $\langle S \rangle_{\rm inf}^4 \lesssim
H_{\rm inf}^2 M_{Pl}^2$, whereas, in SUSY case, we need $\langle S \rangle_{\rm inf} \lesssim M_{Pl}$,
since otherwise the energy density stored in the PQ scalar significantly affects the inflaton dynamics.
}
Alternatively, we may assume that $\zeta$ depends on the inflation field $\phi$. If the inflaton takes a
large field value $\zeta$ may be enhanced during inflation, suppressing the axion quantum fluctuations,
similarly.


 In Fig.~\ref{fig:relaxed-constraint}, we show
the  upper bounds on $H_{\rm inf}$ for $\langle S \rangle_{\rm inf} = \GEV{16}$, $M_{Pl}$, and $15 M_{Pl}$,
which are relaxed with respect to the case of $\langle S \rangle_{\rm inf} = f_a$.
We can see however that super-Planckian values of $\langle S \rangle_{\rm inf}$ are necessary to resolve the
tension between the BICEP2 result and the axion CDM.
There is an interesting possibility that the PQ scalar is the inflaton. Then it takes super-Planckian values of order ${\cal O}(10) M_{Pl}$
during inflation, the axion decay constant  $f_a$ between $\GEV{10}$ and $\GEV{13}$ will be allowed.\footnote{
Here we assume that the kinetic term is not significantly modified at large field values.
In the case of the running kinetic inflation~\cite{Takahashi:2010ky,Nakayama:2010kt,Nakayama:2010sk},
the kinetic term is modified, but the suppression factor is basically same,
as we shall see discuss later.
}

 In the next section, we will show that the tension between the BICEP2 result and the axion CDM can be
resolved without invoking super-Planckian field values for the PQ scalar.

\section{Suppressing isocurvature perturbations by explicit PQ breaking}

\subsection{Basic idea}
Now we give a basic idea to suppress the axion isocurvature perturbations.
 The essential ingredient for our mechanism is the PQ symmetry breaking
enhanced only during inflation. To illustrate our idea,  let us consider a simple set-up,
\bea
{\cal L} &=& \frac{1}{2} (\partial \phi)^2 + \partial S^\dag \partial S - V(\phi, S)
\eea
with
\bea
\label{Vps}
V(\phi,S)&=& \frac{1}{2} m^2 \phi^2 + \frac{\lambda}{4} \left(|S|^2 - f^2 \right)^2 + \delta V_{PQV},
\eea
where $\phi$ is the inflaton, $m$ is the inflaton mass, $\lambda$ is the quartic coupling of $S$,  $f$ determines the
VEV of $|S|$ in the low energy, and $\delta V_{PQ}$ represents the explicit PQ symmetry breaking
terms.   The BICEP2 result \REF{B} is consistent with the chaotic inflation model with a
quadratic potential, but the extension to other inflation models is straightforward.

Let us first consider the following PQ symmetry breaking term,
\bea
\delta V_{PQV} &=& k S^N + {\rm h.c.},
\eea
which breaks the U(1)$_{\rm PQ}$ down to a $Z_N$ subgroup.
Here and in what follows we will take the Planck scale $M_{Pl}\simeq 2.4\times 10^{18}$~GeV to be
unity unless otherwise noted. Suppose that $|S|$ takes a VEV much larger than $f$ during inflation.
This is possible if $S$ is coupled to the inflation sector and acquires a large tachyonic mass.
Then, for sufficiently large $N$, the axion mass receives a sizable contributions,
$\delta m_a^2 \sim k |S|_{\rm inf}^{N-2} $,  during inflation.
If
\bea
\delta m_a^2 \gtrsim H_{\rm inf}^2,
\eea
the axion quantum fluctuations are significantly suppressed at super-horizon
scales. The additional contribution to the axion mass  becomes negligibly small
when $S$ settles down at $|S| = f$ after inflation. Note that $N$ must be very large in order not to spoil the axion
solution to the strong CP problem. For instance, $N > 14$ is required for $f_a = \GEV{12}$~\cite{Carpenter:2009zs}.

Alternatively, we may consider the following interaction,
\bea
\delta V_{PQV} &=& k \phi S^{N'} + {\rm h.c.},
\eea
which depends on the inflaton $\phi$. As  the inflaton takes a super-Planckian field value, the above PQ symmetry breaking
operator is enhanced only during inflation, giving rise to a heavy mass for the axion. This enhancement is efficient
especially in the large-field inflation models.
If the inflaton is stabilized at the origin, this PQ symmetry breaking operator
vanishes in the present vacuum.\footnote{This is indeed the case if we impose a $Z_2$ symmetry on both
$\phi$ and $S$, for $N'$ being an odd integer. } Therefore, in this case, there is no tight lower bound on $N'$.

Thus, one can indeed realize PQ symmetry breaking which becomes relevant only during inflation,
by making use of the dynamics of either the PQ scalar or the inflaton field. We will study the dynamics of the PQ scalar
during inflation in detail in the next subsection.

\subsection{A case in which the PQ symmetry is spontaneously broken during inflation}
Now let us study our scenario in detail.
The enhancement of the PQ symmetry breaking can be realized if  during inflation the saxion acquires a VEV much larger
 than in the present vacuum, or if the  symmetry breaking operators depend on the inflaton field value.
The former possibility requires a relatively flat potential for the saxion.
To this end we consider a SUSY axion models where the saxion has a flat potential lifted only
by SUSY breaking effects and non-renormalizable terms. We however emphasize here that SUSY is not an essential ingredient for our
solution.

The superpotential for the PQ sector can be divided into two parts,
\bea
W =  W_{\rm PQ} + W_{\rm PQV},
\eea
where  $W_{\rm PQ}$ includes the interactions invariant under global U$(1)_{\rm PQ}$, whereas
 $W_{\rm PQV}$ breaks U$(1)_{\rm PQ}$ down to its subgroup $Z_N$.
To be concrete let us consider
\bea
\label{W-PQV}
W_{\rm PQV} = \lambda S^N,
\eea
where $S$ is a PQ superfield developing a vacuum expectation value.
We do not impose U(1)$_R$ symmetry and assume that $\lambda$ is of order unity.\footnote{
This is for simplicity; in the presence of U(1)$_R$ symmetry, the PQ breaking effect can be suppressed
in a certain set-up, and we would need a slightly more involved model to have a sufficiently large PQ breaking.
}   In general, the coefficient of the PQ symmetry
breaking operator can depend on other fields, such as the inflaton.
We shall consider this possibility later.

 The point is that the global U(1)$_{\rm PQ}$ symmetry can be broken to, e.g., a discrete symmetry by quantum gravity effects.
For instance, in string theory, anomalous U$(1)$s which come from gauge symmetries
can be broken to discrete symmetries \cite{Banks:2010zn} by stringy instantons,
which do not exist in the usual gauge theory like QCD \cite{Blumenhagen:2009qh,BerasaluceGonzalez:2011wy}.
%
%


In order to suppress the axion CDM isocurvature perturbations, the saxion must be stabilized
at a large field value during inflation so that the axion mass receives a large contribution
from the PQ-violating operator (\ref{W-PQV}).
On the other hand, after inflation ends,
in order not to spoil the axion solution to the  strong CP problem,
the saxion should settle down at a smaller field value where the axion mass receives a negligible contribution from
the PQ violating operator (\ref{W-PQV}).  For this, the saxion potential needs to be significantly modified during inflation by its coupling
to the inflaton.

Let us first study the condition for the saxion to have a minimum at a large field value.
The relevant terms in the scalar potential of $S$ can be parameterized by\footnote{
We have omitted  interactions with other PQ scalars
assuming that $\arg(S)$ is one of the main components of the axion after inflation,
and is the dominant component during inflation.
}
\bea
\label{V-S}
V = m^2_S |S|^2 - (A_\lambda \lambda S^N + {\rm h.c.}) + \lambda^2 N^2 |S|^{2(N-1)},
\eea
where  soft SUSY breaking terms are included. Note that, during inflation, $m^2_S$ includes
the Hubble-induced mass induced by the coupling to the inflaton. In the present vacuum,
the typical scale for the soft SUSY breaking terms is given by the gravitino mass, $m_{3/2}$. Note that  some of soft SUSY
breaking masses can be loop-suppressed depending on the mediation scheme.
One can find that, if the soft terms satisfy the condition
\bea
\label{condition}
|A_\lambda|^2 > 4(N-1)m^2_S,
\eea
there appears a minimum  at large $S$ where the terms in (\ref{V-S}) dominate the potential.
For instance, if the mass of $S$ is tachyonic, $m_S^2 < 0$, the above condition is satisfied.
Even if the soft mass is non-tachyonic, sufficiently large $A$-term creates a minimum at
a large field value.

Let us examine the scalar potential at large $S$ during inflation.
The interactions with the inflaton field generically induce soft terms as
\bea
\sqrt{|m^2_S|} \sim |A_\lambda| \sim H_{\rm inf},
\eea
for the Hubble parameter $H_{\rm inf}$ larger than $m_{3/2}$.
If the condition (\ref{condition}) is satisfied during inflation, the potential develops
a minimum at
\bea
\label{S-inflation}
|S| \sim \left(\frac{1}{\lambda} \frac{H_{\rm inf}}{M_{Pl}}\right)^{1/(N-2)} M_{Pl}.
\eea
This is the case e.g. if  the mass of $S$ is tachyonic, i.e. the Hubble-induced mass term is negative.
At the local minimum the axion acquires a mass around $H_{\rm inf}$,
\bea
m_a^2 \sim
\lambda A_{\lambda} |S|^{N-2} \sim
H_{\rm inf}^2,
\eea
which highly suppresses  axion quantum fluctuations at super-horizon scales.
Therefore the discrete PQ symmetry provides a simple mechanism to suppress
isocurvature perturbations while protecting the approximate global PQ symmetry against (more) harmful
quantum corrections.

For the suppression mechanism to be viable, it is crucial to make sure that $S$ settles
down to the true vacuum after the inflation is over.
The potential may develop a local minimum at $|S|\sim m^{1/(N-2)}_{3/2}$ due to
the $A$-term after inflation, and then $S$ could be trapped there. This would spoil the axion solution
to the strong CP problem as the axion acquires a too large mass, in general.
The simplest solution to this issue is that the PQ breaking operator
depends on the inflaton field value, and it disappears after inflation. Then, the PQ scalar will settle down
at the true vacuum after inflation without being trapped in the wrong one.
In the case that the PQ breaking term does not depend on the inflaton, we need to consider
the dynamics of the PQ scalar after inflation.
To avoid the trapping in the wrong vacuum, there should be a period during which the condition (\ref{condition})
is violated so that $S$ is pushed toward the true vacuum.
We do not, however,  want the PQ symmetry to be restored, since then the isocurvature perturbations
would be erased irrespective of the above mechanism.

Let us study a concrete axion model to see that  the above conditions can be  indeed realized.
We consider the axion model with
\bea
W_{\rm PQ} &=& \kappa X (S \tilde S - f^2),
\eea
where $S$ and $\tilde S$ have the same  PQ charge with an opposite sign, and $X$ is a PQ singlet.
In the low energy, $S$ and ${\tilde S}$ are stabilized at $\la S \ra \sim \langle {\tilde S} \rangle \sim f$ if
their soft masses are comparable. Thus, the value of $f$ is close to the axion decay constant, $f_a$.
For $H_{\rm inf} < f$, the scalar potential generated by $W_{\rm PQV}$
develops a local minimum at large $S$ along the $F$-flat direction $S\tilde S=f^2$,
if the Hubble-induced mass for $S$  is tachyonic or small enough
compared to the $A$-term so that the condition (\ref{condition}) is satisfied.
The axion then obtains a mass around $H_{\rm inf}$ for $A_\lambda \sim H_{\rm inf}$ at the minimum.
After inflation the Hubble-induced mass for $S$ can be positive, $m_S^2 \sim H^2$, where $H$ denotes the
Hubble parameter.\footnote{
The change of the Hubble-induced mass is possible if the Hubble-induced mass depends on the inflaton
field values in the large-field inflation. Also
this is possible if the inflation ends with the waterfall field as in
the hybrid inflation, since  the Hubble-induced mass can be generated from different couplings in the K\"ahler potential after inflation.}
Then the PQ scalars are stabilized at $S \sim {\tilde S} \sim f$ in the low energy. Note that the U(1)$_{\rm PQ}$ symmetry
can remain  broken even if $S$ acquires a positive Hubble-induced mass after inflation. Let us explain why this is the case.
For $H_{\rm inf} > f$, the PQ symmetry can be broken during inflation if the Hubble-induced mass of $S$ is negative.
After inflation, however,  the PQ symmetry would be restored if both $S$ and ${\tilde S}$ acquire a positive Hubble mass.
On the other hand, the PQ symmetry   remains spontaneously broken if ${\tilde S}$ has a negative Hubble mass after inflation so that
${\tilde S}$ takes a large VEV.\footnote{We assume that ${\tilde S}$ does not participate in the explicit PQ-symmetry breaking.
This may be realized if the PQ symmetry breaking is accompanied with another PQ scalar $S'$ like $S'^M S^N$.
Then the holomorphic nature of the superpotential can forbid or suppress the PQ symmetry breaking due to $S'$. }
After the Hubble parameter becomes sufficiently small, these PQ scalars are stabilized along the
F-flat direction.

It is noted that the saxion starts to oscillate around the true vacuum with a large initial amplitude
after the false vacuum is destabilized.  The saxion oscillation can dominate over the energy density
of the universe after the reheating,  if the saxion is sufficiently long-lived. We have nothing new to
add to the saxion cosmology studied extensively in the literatures
\cite{Choi:1996vz,Kawasaki:2007mk,Nakamura:2008ey,Moroi:2013tea}.
The decay of such saxion reheats the universe again,
and the axion produced from the decay behaves as the dark radiation whereas the SM radiation is produced by the decay into
the Higgs and/or the gauge bosons
\cite{Higaki:2013lra,Jeong:2012np,Higaki:2012ba,Bae:2013bva}.
Interestingly, the presence of dark radiation can solve the tension between the BICEP2 and the Planck
results~\cite{Giusarma:2014zza}.

\begin{figure}[t]
  \begin{center}
    \includegraphics[scale=1.5]{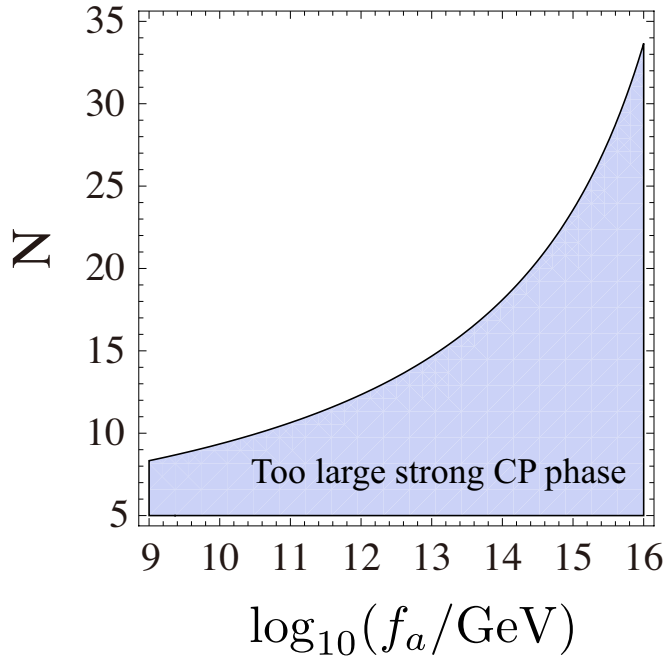}
  \end{center}
  \caption{The lower bound on $N$ as a function of $f_a$. We have set $m_{3/2}=100$TeV,  $\lambda = 1$ and $A_\lambda = m_{3/2}$.}
  \label{fig:N}
\end{figure}

Lastly let us examine how large $N$ should be in order not to spoil the axion solution to the strong CP problem.
In general, the PQ breaking term contributes to an additional CP phase.  The contribution
 becomes sufficiently small when the axion mass is dominated by the QCD
instanton contribution,
\bea
\Delta m^2_a \sim A_\lambda \lambda N^2 f^{N-2}_a \,<\, 10^{-11}\,(m^{\rm QCD}_a)^2,
\eea
where $m^{\rm QCD}_a$ is given by \EQ{maQCD}, and  $\Delta m^2_a$ represents the contribution of
the PQ breaking term to the axion mass in the true vacuum.
This condition puts a lower bound on $N$. We show it in Fig.~\ref{fig:N} for the case with $\lambda\sim 1$ and
$A_\lambda \sim m_{3/2}$. For instance, $N > 12$ is necessary for $f_a = \GEV{12}$.
If PQ symmetry breaking terms are present only during inflation, there is no such lower bound.

%
%
%
%

\subsection{A case in which the PQ symmetry is spontaneously broken after inflation}
If the PQ symmetry is restored during and/or after inflation, and if it becomes spontaneously
broken after inflation, there is no isocurvature constraint. In this case, however,
axionic cosmic strings are produced at the phase transition, whereas domain walls are
generated at the QCD phase transition. The domain wall number $N_{\rm DW}$
must be unity to  avoid the overclosure of the Universe; in this case, the axions are copiously
produced by domain wall annihilations, and
the right amount of axion dark matter is generated for $f_a \approx \GEV{10}$~\cite{Hiramatsu:2012gg}.

If there is an additional PQ symmetry breaking term enhanced during inflation,
we can actually relax the above constraint on the domain wall number and the decay constant.
To simplify our argument, let us assume that the explicit PQ symmetry breaking is proportional to $S$,
and that the coefficient depends on another scalar field $\psi$. The PQ symmetry breaking
can be temporarily enhanced, if $\psi$ takes a large VEV until some time
after the PQ symmetry breaking, and it settles at the origin in the low energy.
The point is that, after the PQ symmetry breaking,  the explicit breaking term induces domain walls
attached to the axionic strings. Then,  if there is only unique vacua, i.e., the  domain wall number associated
with the PQ symmetry breaking is unity, strings and domain walls disappear when the domain wall energy
exceeds that of axionic strings. (Note that the domain wall number $N_{\rm DW}$ for the QCD anomaly has nothing
to do with the domain wall number associated with the extra PQ symmetry breaking.)
When domain walls and cosmic strings annihilate, axions are produced.
Such axions will be diluted by cosmic expansion and/or decay into the SM particles as their mass remains heavy
until the explicit PQ symmetry breaking becomes suppressed.
Interestingly, the axion field value is set to be a value which minimizes
the PQ symmetry breaking after the domain wall annihilation. Thus, the large decay constant as well as $N_{\rm DW} > 1$
can be allowed in the presence of such temporarily enhanced PQ symmetry breaking.

\section{Discussion and Conclusions}

We have seen that, if the PQ scalar takes super-Planckian values during inflation,
there is an allowed region for $f_a$ where the BICEP2 results become consistent with the axion CDM.
Let us here briefly  discuss one inflation model with the PQ scalar identified with the inflaton.
In non-SUSY case, the PQ scalar can be stabilized by the balance between the negative mass and the quartic coupling
as in \EQ{Vps}.
Then, it is possible to realize the quadratic chaotic inflation model with the PQ scalar, if the kinetic
term of the PQ scalar is significantly modified at large field values, based on the running kinetic
inflation~\cite{Takahashi:2010ky,Nakayama:2010kt,Nakayama:2010sk}. For instance, we can consider
\bea
{\cal L} &=&|\partial S|^2 + \xi (\partial |S|^2)^2 - \frac{\lambda}{4} \left(|S|^2-f^2 \right)^2,
\eea
where $\xi \gg 1$ denotes the coefficient for the running kinetic term, and $\lambda$ is
the quartic coupling. The large value of $\xi$ can be understood by imposing a shift symmetry
on $|S|^2$: $|S|^2 \to |S|^2 + C$, where $C$ is a real constant. Then the $\xi$-term respects the
symmetry, while the other terms explicitly break the shift symmetry.\footnote{
In other words, the ordinary kinetic term and the potential term are relatively suppressed
as they explicitly break the shift symmetry.
} Note that the shift symmetry is
consistent with the PQ symmetry, as it is the radial component of $S$ that transforms under the
shift symmetry. Let us denote the radial component of $S$ as $\sigma = |S|$.
At large field values $\sigma \gg 1/\sqrt{\xi}$,
the canonically normalized field is ${\hat \sigma}\sim \sqrt{\xi} \sigma^2$, and the scalar potential becomes
the quadratic one in terms of $\sigma$. The quadratic chaotic inflation can be realized by the PQ scalar.
In this case, $\langle {\hat S} \rangle_{\rm inf}$
is of order $15 M_{Pl}$, and
one can see from Fig.\ref{fig:relaxed-constraint} that
there is a region between $f_a \simeq \GEV{10}$ and $\GEV{13}$ where the isocurvature
constraint becomes consistent with the BICEP2 result. Thus, one interesting way to evade the isocurvature bound
is to identify the PQ scalar with the inflaton.

In this letter, we have proposed a simple mechanism in which the axion acquires a heavy mass
during inflation, leading to the suppression of the axion CDM isocurvature perturbations. The
point is that the U(1)$_{PQ}$ symmetry is explicitly broken down to its discrete subgroup,  $Z_N$.
If the PQ-breaking operators are significant during inflation,  the axion can acquire a sufficiently heavy
mass, suppressing the isocurvature perturbations. There are two ways to accomplish this.
One is that the PQ-breaking operators depend on the inflaton field value.
If the inflaton takes a large-field value during inflation, the PQ-breaking operator becomes significant,
while it becomes much less prominent if the inflaton is stabilized at smaller field values after inflation.
This 
possibility nicely fits with the
large field inflation suggested by the BICEP2 result \REF{B}. The other is that the saxion field
value changes significantly during and after inflation. Then, for sufficiently large power of the PQ symmetry
breaking operator, it is possible to realize the heavy axion mass during inflation, while keeping the
axion solution to the strong CP problem.
%
The lower bound on the power $N$ reads $N > 12$ for $f_a = \GEV{12}$ and the gravitino mass of order $100$\,TeV.
We have discussed a concrete PQ scalar stabilization to show that it is possible that the saxion
is fixed at a large field value during inflation, while it settles down at the true minimum located
at a smaller field value, without restoring the PQ symmetry.
%

Toward a UV completion of our scenario based on the string theory, we may consider non-perturbative effects,
\bea
W \supset (\phi -\phi_0)^n e^{\pm{\cal A}}, ~~~{\rm or}~~~ K \supset  (\phi^{\dag} -\phi_0^{\dag})^m e^{\pm{\cal A}} + {\rm h.c.},
\eea
where the exponentials $\propto e^{\pm{\cal A}}$ break a $U(1)_{\rm PQ}$ symmetry down to a discrete one. Here
${\cal A}$ is a (linear combination of) string theoretic axion multiplet, and $\phi$ is the inflaton.
If the inflaton $\phi$ develops non-zero expectation value during the inflation (and $\phi=\phi_0$ in the true vacuum),
such an axion obtains a large mass, which may suppress the isocurvature perturbations.
Similar terms may also help the overshooting problem of moduli during inflation,
producing the high potential barrier against decompactification \cite{Kallosh:2004yh}\footnote{
The so-called Kallosh-Linde model can give a large mass to the relevant axion multiplet:
$W \supset (\phi -\phi_0)^n (W_0 + Ae^{\pm a{\cal A}}+ Be^{\pm b{\cal A}})$.
See also \cite{He:2010uk}.
}, even without the coupling to the inflaton in the superpotential \cite{Abe:2005rx}.

In nature there may be other kinds of pseudo Nambu-Goldstone bosons such as axion-like particles,
and the isocurvature perturbation bound can be similarly applied to them. The tension between
the isocurvature perturbations and the high-scale inflation can be solved by our mechanism.
That it to say, we can add a symmetry breaking operator which becomes relevant only during inflation.
We can do so by either introducing an inflaton field charged under the symmetry or by assuming that
the radial component significantly evolves during and after inflation. In particular, our mechanism
can relax the isocurvature bound on the $7$\,keV axion dark matter proposed by the present authors~\cite{Higaki:2014zua}
to explain the recently found $3.5$\,keV X-ray line~\cite{Bulbul:2014sua,Boyarsky:2014jta}.

{\it Note added:} Isocurvature constraints on the QCD axion and axion like particles
were considered also in Refs.~\cite{Marsh:2014qoa,Visinelli:2014twa} soon after the BICEP2
announcement. 

{\it Note added 2:} After submission of this paper we noticed that some of our arguments has an
overlap with Ref.~\cite{Dine:2004cq} where it was pointed out
that the axion isocurvature perturbations can be suppressed if the PQ symmetry is badly broken 
during inflation and the alignment of the axion by the PQ symmetry breaking can eliminate domain walls. 

\section*{Acknowledgment}
This work was supported by Grant-in-Aid for  Scientific Research on Innovative
Areas (No.24111702, No. 21111006, and No.23104008) [FT], Scientific Research (A)
(No. 22244030 and No.21244033) [FT], and JSPS Grant-in-Aid for Young Scientists (B)
(No. 24740135 [FT] and No. 25800169 [TH]), and Inoue Foundation for Science [FT].
This work was also supported by World Premier International Center Initiative
(WPI Program), MEXT, Japan [FT].

\end{document}